\documentclass{IEEEtran}
\usepackage{amssymb,color,subfigure}
\usepackage{graphicx}
\usepackage{xspace,stackrel,hyperref}
\usepackage{wrapfig}
\usepackage{tikz,pgfplots}
\usetikzlibrary{external} 
\usepackage{autonum}

\usetikzlibrary{shapes,positioning}
\usetikzlibrary{narrow}

\usepackage{dsfont}

\definecolor{mycolor4}{RGB}{230,97,1}
\definecolor{mycolor2}{RGB}{178,171,210}
\definecolor{mycolor3}{RGB}{253,184,99}
\definecolor{mycolor1}{RGB}{94,60,153}

\makeatletter 
\pretocmd\@bibitem{\color{black}\csname keycolor#1\endcsname}{}{\fail}
\newcommand\citecolor[1]{\@namedef{keycolor#1}{\color{blue}}}
\makeatother

\DeclareMathAlphabet{\mathcal}{OMS}{cmsy}{m}{n}

\newcommand{\mc}{\mathcal}

\newcommand{\1}{\mathds{1}}
\newcommand{\E}{\mc{E}}

\newcommand{\R}{\mathds{R}}

\DeclareMathOperator*{\esssup}{ess\,sup}
\newcommand{\one}{\mathds{1}}

\newcommand{\map}[3]{#1: #2 \rightarrow #3}

\newcommand{\norm}[1]{\left\lVert#1\right\rVert}
\renewcommand{\d}{\text{d}}

\DeclareMathOperator*{\argmin}{arg\,min}

\newcommand{\retainlabel}[1]{\label{#1}\sbox0{\ref{#1}}}

\definecolor{mycolor1}{RGB}{230,97,1}
\definecolor{mycolor2}{RGB}{178,171,210}
\definecolor{mycolor3}{RGB}{253,184,99}
\definecolor{mycolor4}{RGB}{94,60,153}
\definecolor{mycolor5}{rgb}{0,0,0}

\newtheorem{assumption}{Assumption}

\newtheorem{theorem}{Theorem}
\newtheorem{proposition}{Proposition}

\newtheorem{definition}{Definition}

\newtheorem{remark}{Remark}

\newtheorem{example}{Example}

\title{\LARGE\bf The Strong Integral Input-to-State Stability Property in \\ Dynamical Flow Networks}
\author{Gustav Nilsson and Samuel Coogan\thanks{
G. Nilsson is with the School of Architecture, Civil and Environmental Engineering, École Polytechnique Fédérale de Lausanne (EPFL), 1015 Lausanne, Switzerland. {\tt\small gustav.nilsson@epfl.ch}.
S. Coogan is with the School of Electrical and Computer Engineering, Georgia Institute of Technology, Atlanta, 30332, USA. {\tt\small sam.coogan@gatech.edu}. S.~ Coogan is also with the School of Civil and Environmental Engineering, Georgia Institute of Technology.

This work was supported by the National Science Foundation under grant \#1749357 and the Air Force Office of Scientific Research under award FA9550-19-1-0015. A preliminary version of some of the results in this paper was presented in~\cite{nilsson2021strongiiss}.
}}

\begin{document}

\maketitle

\begin{abstract}
Dynamical flow networks serve as macroscopic models for, e.g., transportation networks, queuing networks, and distribution networks. While the flow dynamics in such networks follow the conservation of mass on the links, the outflow from each link is often non-linear due to, e.g., flow capacity constraints and simultaneous service rate constraints. Such non-linear constraints imply a limit on the magnitude of exogenous inflow that is able to be accommodated by the network before it becomes overloaded and its state trajectory diverges. This paper shows how the Strong integral Input-to-State Stability (Strong iISS) property allows for quantifying the effects of the exogenous inflow on the flow dynamics. The Strong iISS property enables a unified stability analysis of classes of dynamical flow networks that were only partly analyzable before, such as networks with cycles, multi-commodity flow networks and networks with non-monotone flow dynamics.  We present sufficient conditions on the maximum magnitude of exogenous inflow to guarantee input-to-state stability for a dynamical flow network, and we also present cases when this sufficient condition is necessary. The conditions are exemplified on a few existing dynamical flow network models, specifically, fluid queuing models with time-varying exogenous inflows and multi-commodity flow models.
\end{abstract}

\begin{IEEEkeywords}
dynamical flow networks, input-to-state stability, transportation networks, queuing networks
\end{IEEEkeywords}

\section{Introduction}
Dynamical flow networks serve as macroscopic models for physical network flows such as transportation networks~\cite{lovisari2014stability, grandinetti2015, coogan2015compart} as well as non-physical processing networks such as queuing systems~\cite{dai2020processing}. One frequent common component for those networks is a limitation on the magnitude of exogenous inflow the networks can accommodate due to, e.g., maximum flow capacity on the roads or maximum processing capacity at a server. The dynamics of such networks is often non-linear, both due to the physical flow dynamics itself and saturation in the service rates. Thus, classical techniques such as linear system analysis are not enough to analyze these systems' stability properties.

In many of the aforementioned applications, the goal is to keep link densities or queues bounded. It is generally observed that this is possible as long as the exogenous inflow to the network stays below a certain threshold. The purpose of this paper is to formalize this observation as necessary and sufficient conditions for stability for a large class of dynamic flow networks. 
The Strong Integral Input-to-State property (Strong iISS) was introduced in~\cite{strongiISS} for general dynamical systems to combine integral input-to-state stability (iISS) with input-to-state stability (ISS) for small inputs and to determine when the input is small enough to guarantee the latter. Although those two properties align naturally with the expected behavior of dynamical flow networks, to the best of the authors' knowledge, the Strong iISS property has not been exploited for studying dynamical flow networks in a general setting with features such as cycles, multi-commodity flows, and non-monotone flow dynamics. In the context of dynamic flow networks, apart from having the desired property of guaranteeing stability when the exogenous inflow to the network is lower than a certain threshold, the Strong iISS property also imposes that if the exogenous inflow becomes zero at a certain time, the total mass in the network will also eventually converge to zero. For many applications, this is an essential property. For example, in transportation networks, a correct traffic signal control solution should allow for the vehicles to eventually leave the network.

Existing literature on stability of dynamic flow networks often focuses on the class of networks with dynamics that exhibit a monotonicity property whereby state trajectories maintain a partial order on the state space \cite{como2013robustI, como2017review, coogan2019contractive}, or a related mixed-monotonicity property \cite{coogan2016stability}. In these cases, powerful results from monotone systems theory and contraction theory provide sufficient conditions for stability and, e.g., constructive methods for obtaining Lyapunov functions. 

However, as noticed in~\cite{nilsson2014multicommodity}, when extending  flow network models to multi-commodity flows, the monotonicity property is usually lost. Another example when the system's monotonicity property is lost is when a feedback controller can serve more than one queue simultaneously, and the service is split in proportion to the demand in all queues that are served simultaneously~\cite{savla2014}. This situation is common in many applications, such as when controlling traffic signals in a transportation network. Other methods that have been proposed for considering specific classes of networks include using passivity theory~\cite{bianchin2020}
or constructing specific entropy-like Lyapunov functions~\cite{nilsson2015entropy}, but a general framework for considering non-monotone dynamic flow networks remains elusive.

In this paper, we propose Strong iISS as another important tool, alongside monotone systems theory and contraction theory, for studying a large class of dynamic flow networks. The stability analysis in this paper partly relies on a special variant of sum-separable Lyapunov functions similar to those that have previously been combined with monotonicity properties, e.g.,~\cite{coogan2019contractive, rantzer2015}. In particular, the Lyapunov function is based on a transformation involving the inverse of the routing matrix for the network. This transformation has previously been utilized to obtain monotonicity properties for tree-like flow networks in~\cite{schmitt2018optimal}. 

Preliminary results on using Strong iISS to study dynamical flow networks appeared in~\cite{nilsson2021strongiiss}. In this paper, we extend those results by considering both bounded and unbounded flow functions. We also show that, in the case of bounded flow functions, it is always optimal to normalize the stability condition by the capacity vector. Furthermore, we extend the stability analysis to network flow dynamics described by differential inclusions. This dynamic description is needed when there is a possibility that a link with no mass present can receive service, something that is common in, e.g., traffic network applications where several links may belong to the same service phase. We also extend the examples beyond those presented in~\cite{nilsson2021strongiiss}.

The rest of the paper is organized as follows: The remainder of this section is devoted to introducing some basic notation that will be used throughout the paper. In Section~\ref{sec:model}, we present the dynamical flow network model, together with a few general model assumptions. We also show that under those mild assumptions, dynamical flow networks are always integral input-to-state stable. In Section~\ref{sec:stability} we show that a dynamical flow network is Strong iISS. We present a sufficient condition on the exogenous inflow for the dynamical flow networks to be input-to-state stable and show how this condition can be tightened when all the flow functions are bounded. We also present a case when the ISS condition is tight for a local network, and an alternative bound on the growth rates of the state, that differs from the standard Strong iISS bound. In the following section, Section~\ref{sec:diffinc}, we extend the analysis to network flow dynamics where the outflow is described through differential inclusion. In Section~\ref{sec:examples}, we illustrate how the stability theory can be applied to existing models for dynamical flow networks, namely dynamical networks with time-varying exogenous inflows and multi commodity flow networks. The paper is concluded with some ideas for future research.

\subsection{Notation}
We let $\R_+$ denote the non-negative reals. For a finite set $\mc A$, $\R_+^{\mc A}$ denote the set of non-negative vectors indexed by $\mc A$. Unless stated otherwise,  for a vector $x \in \R^n$, we let $\norm{x}$ denote the $\ell_1$ norm. For vectors, all inequalities apply element-wise. For vectors $w,x \in \R^n_+$ such that $w >0$, we introduce the weighted $\ell_1$-norm as $\norm{x}_w = w^T x$. The all-one vector is denoted by $\one$. A function $\map{\mu(x)}{\R_+}{\R_+}$ is said to be of class $\mc K_\infty$ if it is strictly increasing, $\mu(0) = 0$, and $\lim_{x \rightarrow +\infty} \mu(x) = +\infty$. A function $\map{\beta(x,t)}{\R_+\times \R_+}{\R_+}$ is said be of class $\mc K \mc L$ if $\beta(0,t) = 0$ for all $t$, it is strictly increasing in $x$ for each fixed $t$, and it is decreasing in $t$  for each fixed $x$ and $\lim_{t\rightarrow +\infty} \beta(x,t) \rightarrow 0$.

\section{Model}\label{sec:model}
We model a dynamical flow network as a directed multi-graph, i.e., a directed graph where there can exist multiple parallel links between two nodes, $\mc G = (\mc V, \mc E)$, where $\mc V$ is the set of nodes and $\mc E$ is the multiset of links. We will assume that the graph has no self loops. For a link $e = (i,j) \in \mc E$, we let $\tau(e)$ denote the tail of the link, i.e., $\tau(e) = i$, and $\sigma(e)$ the head of the link, i.e., $\sigma(e) = i$. Moreover, we let $\mc E_v$ denote the subset of incoming links to node $v \in \mc V$, formally $\mc E_v = \{i \in \mc E \mid \sigma(i) = v\} \subset \mc E$. 

In the flow network, mass flows along the links $\mc E$. Therefore, the network's state $x \in \mc X = \R_+^{\mc E}$ is the vector of masses on all the links in the network and $\mc X$ will be the short notation for the state space. We let  $\lambda(t) \in  \R_+^{\mc E}$ denote the vector of possible time varying exogenous inflow to the links, and for links that can have no exogenous inflow, the corresponding elements in $\lambda(t)$ will be zero.

To model the mass propagation through the network, we introduce the routing matrix $R \in \R^{\mc E \times \mc E}$. An element $R_ij$ in the routing matrix, tells the fraction of outflow from link $i$ that will proceed to link $j$. Hence, $0 \leq R_{ij} \leq 1$. Moreover, due to conservation of mass it must hold that for all $i \in \mc E$, $\sum_j R_{ij} \leq 1$ where $1- \sum_j R_{ij}$ is the fraction of mass that leaves the network after flowing out from link $i$. Since the routing matrix has to obey the network topology, $R_{ij} > 0$ only if $\sigma(i) = \tau(j)$.

Throughout the paper, we will make the assumption that the network is outflow connected, i.e., from every link in the network it is possible to find a path to a link where a fraction of the mass can leave the network.

\begin{assumption}\label{ass:routing}
The routing matrix $R$ is assumed to be out-flow connected, i.e., for every link $i \in \mc E$ there exists a path to a link $j \in \mc E$ such that $\sum_k R_{jk} < 1$, where the existence of a path to the link $j$ can be equivalently expressed as that there exists an $\ell > 0$ such that $(R^\ell)_{ij} > 0$.
\end{assumption}
\medskip

Assumption~\ref{ass:routing} implies that the spectral radius of the routing matrix is less than one, and hence the matrix $(I-R^T)$ is invertible~\cite{como2016local}. Its inverse can be computed through
\begin{equation}
(I-R^T)^{-1} = \sum_{k \geq 0} (R^T)^k = I + R^T + (R^T)^2 + \ldots\,.   
\end{equation}

The outflow from each link is controlled by a state-dependent Lipschitz-continuous outflow function, which we denote $f_i(x) : \R_+^{\mc E} \rightarrow \R_+$ for every link $i \in \mc E$. We let $f(x)$ denote the vector of all flow functions, i.e., $f(x) = (f_i(x))_{i\in \mc E}$. In this setting we allow the outflow function to depend on the link's own state, which is common in applications where the outflow only depends on the mass transportation dynamics on the link itself. But we also allow for the outflow to be dependent on the state of the neighboring links or the whole network, something that is common in, e.g., queueing theoretic applications where the service has to be split among different links.

We will make the following assumption about the flow functions.
\begin{assumption}\label{ass:flow}
We assume that the flow functions $f_i(x)$ are always non-negative and such that $f_i(x) = 0$ if and only if $x_i = 0$.
\end{assumption}
\medskip

The ``if'' part of the assumption ensures that the mass always stays non-negative. The ``only if'' part ensures that the flow functions are work-conservative, i.e., if there is a mass present on one link, at least some of it will flow out from the link. In Section~\ref{sec:diffinc}, we will relax the ``if'' part of this assumption to adopt a broader set of outflow functions.

\begin{remark}
In contrast to some related work on dynamical flow networks, e.g.,~\cite{como2017review}, we impose no monotonicity assumptions on the flow functions.
\end{remark}
\medskip

The dynamics then follows from the conservation of mass: the change of mass on each link equals the inflow from the upstream links combined with potential exogenous inflow minus the outflow from the link itself. That is, 
\begin{equation}
    \dot{x}_i = \lambda_i +\sum_{j \in \mc E} R_{ji} f_j(x) - f_i(x)  \,, \quad \forall i \in \mc E\,,
\end{equation}
which is expressed in vector form as
\begin{equation}
\dot{x} = \lambda - (I-R^T)f(x) \,. \label{eq:generaldynamics}
\end{equation}

In many applications, the desired stability properties of the network flow dynamics~\eqref{eq:generaldynamics} under Assumptions~\ref{ass:routing} and~\ref{ass:flow} are the following:
\begin{enumerate}
\item In the case of unbounded flow functions, the state should always stay bounded when the exogenous inflows are bounded. If the flow functions $f(x)$ are bounded from above, then for small enough $\lambda$, the state $x$ should stay bounded. 
\item If the exogenous inflow $\lambda$ becomes  zero, then the state $x$ should eventually become zero as well.
\end{enumerate}

The first property is often referred to as stability of dynamical flow networks. The second property ensures that all the mass in the network will eventually leave the network, something that is desirable in, e.g., transportation network applications. 

In the next section, we will see how these desired properties fits well with the Strong integral Input-to-State Stability (Strong iISS) property and show that the dynamical flow network dynamics~\eqref{eq:generaldynamics} is Strong iISS.

\section{Strong iISS Property for Dynamical Flow Networks} \label{sec:stability}

In the first part of this section, we show that dynamical flow networks satisfying Assumptions~\ref{ass:routing} and \ref{ass:flow} are Strong iISS, which includes a sufficient condition on the exogenous inflows for the system to be ISS. We then show that this sufficient condition can be tightened by normalizing by the flow capacities when all the flow functions are bounded. While it is possible to normalize the condition by any positive vector, we show that the capacity vector is the optimal choice. We then show a special case when the sufficient condition is also necessary. In the last part of this section, we present a alternative bound to the standard Strong iISS bound on the state's growth rate.

\subsection{Dynamical Flow Networks are Strong iISS}
Let us first recall the definition of Strong iISS from~\cite{strongiISS}.

\begin{definition}[Strong iISS, {\cite[Def. 2]{strongiISS}}]\label{def:StrongiISS}
A dynamical system $\dot{x} = f(x, u)$ is said to be Strong iISS if there exists a function $\beta \in \mc K \mc L$, and functions $\mu_1, \mu_2, \mu \in \mc K_\infty$ such that for all $x(0)$ and all $t \geq 0$ it holds that
\begin{enumerate}
    \item \begin{equation} \norm{x(t)}  \leq \beta(\norm{x(0)}, t) + \mu_1 \left( \int_0^t \mu_2(\norm{u(s)})\d s \right) \,,\end{equation}
    \item moreover, there exists a constant $U > 0$, such that when
    \begin{equation}
    \esssup_{t\geq 0} \norm{u(t)} < U \,,
    \end{equation}
    it holds that
    \begin{equation}
    \norm{x(t)} \leq \beta(\norm{x(0)}, t) + \mu\left(    \esssup_{t\geq 0} \norm{u(t)} \right ) \,.
    \end{equation}

\end{enumerate}
\end{definition}

\medskip

The first part of the definition states that the size of the state is limited by the integral of the input and is the classical integral Input-to-State Stability (iISS) requirement~\cite{sontag98integral}. The second part of the definition implies that the state is always bounded when the input is small enough. In the case when the definition is valid for any choice of $U$, the system has the classical Input-to-State Stability (ISS) property. It should be noted that when there exists a time $t' \geq 0$ such that the input $u(t)= 0$ for all $t \geq t' \geq 0$, then the state will eventually converge to zero. Hence, the second part of Definition~\ref{def:StrongiISS} captures the desired stability properties mentioned in Section~\ref{sec:model}.

Next, we will prove that dynamical flow networks are Strong iISS.

\begin{theorem}\label{thm:cyclic}
A dynamical flow network~\eqref{eq:generaldynamics} satisfying Assumption~\ref{ass:routing} and~\ref{ass:flow} is iISS, i.e., there exist $\beta \in \mc K \mc L$ and $\mu_1, \mu_2 \in \mc K_\infty$ such that
\begin{multline}
    \norm{(I-R^T)^{-1}x(t)} \leq \beta(\norm{(I-R^T)^{-1}x(0)}, t)  + \\ \mu_1\left( \int_0^t \mu_2(\norm{(I-R^T)^{-1}\lambda(s)}) \d s \right) \,, \label{eq:iISS}
\end{multline}
for all $t \geq 0$.

Moreover, let $a(t) = (I-R^T)^{-1} \lambda(t)$. If
\begin{equation}
\esssup_{t\geq 0} \sum_{i \in \mc E} a_i(t) < \liminf_{\norm{x}\rightarrow +\infty} \sum_{i \in \mc E} f_i(x) \,,  \label{eq:condcyclic}
\end{equation}
then the dynamical flow network is also ISS, i.e., there exists functions $\beta \in \mc K \mc L$ and $\mu \in \mc K_\infty$ such that the solution to the dynamical flow~\eqref{eq:generaldynamics} satisfies
\begin{multline}
    \norm{(I-R^T)^{-1}x(t)} \leq \beta(\norm{(I-R^T)^{-1} x(0)}, t)  + \\ \mu\left(\esssup_{t \geq 0} \norm{(I-R^T)^{-1}\lambda(t)}\right) \,. \label{eq:ISSunbounded}
\end{multline}
Hence, the dynamical flow network is Strong iISS.
\end{theorem}

\medskip

\begin{IEEEproof}
Introduce the Lyapunov candidate 
\begin{equation}
V = \1^T(I-R)^{-1}x \,. \label{eq:lyapcand}
\end{equation}
Since $x \geq 0$ and $(I-R^T)^{-1} = \sum_{k\geq 0} (R^T)^k$, the Lyapunov candidate~\eqref{eq:lyapcand} satisfies $V(x) \geq 0$ and $V(x) = 0$ if and only if $x=0$. Moreover,
\begin{multline}
\frac{\d V}{\d t} = \frac{\partial V}{\partial x} f(x) = \1^T (I-R^T)^{-1} (\lambda - (I -R^T)f(x)) \\ = \sum_{i \in \mc E} a_i(t) - \sum_{i \in \mc E} f_i(x) \,. \label{eq:dVdt}
\end{multline}

Let $\gamma(x) = x$ and $W(x) = \sum_{i \in \mc E} f_i(x)$. Clearly, $\gamma \in \mc K_\infty$, and
\begin{equation}
\gamma(\norm{(I-R^T)^{-1}\lambda(t)}) = \sum_{i \in \mc E} a_i(t) \,.
\end{equation}
Eq.~\eqref{eq:dVdt} can be rewritten
\begin{equation}
\frac{\d V}{\d t} = \frac{\partial V}{\partial x} f(x) = -W(x) + \gamma(\norm{(I-R^T)^{-1}\lambda(t)}) \,.
\end{equation}
Due to Assumption~\ref{ass:flow}, $W(x)$ is positive definite. Now, applying~\cite[Theorem 1]{strongiISS} gives the bound.
\end{IEEEproof}

\subsection{The Strong iISS Property for Bounded Flow Functions}

In the case when all the flow functions are bounded such that outflow capacity from each link is given by $c_i = \sup_{x\geq 0} f_i(x)$, it is possible to state a tighter bound for when the dynamical flow network is ISS. Let $c = (c_i)_{i \in \E}$ be the vector of all capacities, $\bar{c} = (1/c_i)_{i \in \mc E}$ and $C = \text{diag}(c)$. Moreover, for all links $i \in \mc E$, introduce the normalized flow functions
\begin{equation}
\tilde{f}_i(x) = \frac{f(x)}{c_i} \,,
\end{equation}
and let $\tilde{f}(x) = (\tilde{f}_i(x))_{i\in \mc E}$ be the vector of all flow functions. The dynamics~\eqref{eq:generaldynamics} can then equivalently be written as 
\begin{equation}
\dot{x} = \lambda - (I-R^T)C\tilde{f}(x) \,. \label{eq:dynnorm}
\end{equation}

\begin{theorem}\label{thm:cyclicbounded}
Consider the dynamical flow network with normalized flow functions~\eqref{eq:dynnorm}. Assume that Assumptions~\ref{ass:routing} and~\ref{ass:flow} are satisfied and let $a(t) = (I-R^T)^{-1} \lambda(t)$. Then, if 
\begin{equation}
\esssup_{t\geq 0} \sum_{i \in \mc E} \frac{a_i(t)}{c_i} < \liminf_{\norm{x}\rightarrow +\infty} \sum_{i \in \mc E} f_i(x) \,,  \label{eq:condnormed}
\end{equation}
there exists functions $\beta \in \mc K \mc L$ and $\mu \in \mc K_\infty$ such that the solution to the dynamical flow~\eqref{eq:dynnorm} satisfies
\begin{multline}
    \norm{(I-R^T)^{-1}x(t)}_{\bar{c}} \leq \beta(\norm{(I-R^T)^{-1} x(0)}_{\bar{c}}, t) + \\ \mu\left(\esssup_{t \geq 0} \norm{(I-R^T)^{-1}\lambda(t)}_{\bar{c}}\right) \,, \label{eq:ISSnormed}
\end{multline} 
i.e., the dynamical flow network is ISS.
\end{theorem}

\begin{IEEEproof}
The proof follows the same way as the proof for Theorem~\ref{thm:cyclic}, but with the Lyapunov candidate 
\begin{equation}
V(x) = \1^T C^{-1} (I-R^T)^{-1}x \,.
\end{equation}
\end{IEEEproof}

Although, the bound in~\eqref{eq:ISSnormed} in Theorem~\ref{thm:cyclicbounded} is different from the bound~\eqref{eq:condcyclic} in Theorem~\ref{thm:cyclic}, the sufficient condition in Theorem~\ref{thm:cyclicbounded} can actually be less conservative than its counterpart in Theorem~\ref{thm:cyclic}, as the following example shows.

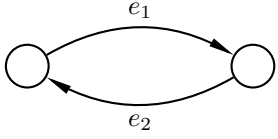
\begin{figure}
\centering
\begin{tikzpicture}[>=narrow,thick]
\node[draw, circle] (a) at (0,0){\phantom{$v$}};
\node[draw, circle] (b) at (3,0){\phantom{$v$}};
\draw[->] (a) to[bend left] node[above] {$e_1$} (b);
\draw[->] (b) to[bend left] node[below] {$e_2$} (a);
\end{tikzpicture}
\caption{The two node network used in Example~\ref{ex:conditiontight} to illustrate a case when the bound for ISS in Theorem~\ref{thm:cyclic} is tight.}
\label{fig:twonodes}
\end{figure}

\begin{example}\label{ex:conditiontight}
Consider the two-link, two-node network in Figure~\ref{fig:twonodes}. Let $f_1(x_1) = c_1(1-\exp(-x_1))$ and   $f_2 (x_2) = c_2(1-\exp(-x_2))$ with $c_1 = 1$ and $c_2 \gg 1$. Let the routing matrix $R$ be such that $R_{1,2} = 0.9$ and $R_{2,1} = 1$. Then $a_1 = 10\lambda_1 + 10\lambda_2$ and $a_2 = 9\lambda_1 + 10\lambda_2$.

Using the condition in Theorem~\ref{thm:cyclic}, the sufficient condition for ISS reads 
\begin{equation}
19 \lambda_1 + 20 \lambda_2 < 1 \,,
\end{equation}
but by utilizing the fact that the functions are bounded and the condition in Theorem~\ref{thm:cyclicbounded} instead, the sufficient condition becomes 
\begin{equation}
\frac{\lambda_1 + \lambda_2}{c_1} + \frac{\lambda_1 + \lambda_2}{c_2} < 1 \,
\end{equation}
which allows for larger exogenous inflows when $c_2$ is large. In fact, when $c_2 \rightarrow +\infty$, the condition above becomes 
\begin{equation}
\lambda_1 + \lambda_2 < 1 
\end{equation}
which is arbitrary close to the necessary condition that $\lambda_1 + \lambda_2 \leq 1$.

\end{example}

\medskip

The previous example raises the question of whether, in the case of bounded flow functions, another choice of normalizing vector instead of the capacity vector $c$ can yield a more relaxed sufficient condition. To answer this question, we start by observing that for the case when all the flow functions are such that
\begin{itemize}
\item the flow only depends on the mass on the link itself, i.e., $\frac{\partial}{\partial x_j} f_i(x) =0$ for all $j \neq i \in \mc E$, and
\item the flow attains it maximum in its limit, i.e., $\lim_{x_i\rightarrow +\infty} f_i(x) = c_i$ for all $i \in \mc E$,
\end{itemize}
the condition~\eqref{eq:condnormed} in Theorem~\ref{thm:cyclicbounded} reads
\begin{equation}
\esssup_{t\geq 0} \sum_{i \in \mc E} \frac{a_i(t)}{c_i} \leq \min_{i \in \mc E} c_i \,.
\end{equation}

The following proposition shows that in this case, the capacity is the optimal choice of normalizing vector.
\begin{proposition}
Given vectors $a \in \R^{\mc E}$ and $c \in \R_+^{\mc E}$ where $c > 0$,  suppose that there exists a strictly positive vector $b \in \R_+^{\mc E}$, $b>0$ such that
\begin{equation}
\sum_{i \in \mc E} \frac{a_i}{b_i} \leq  \min_{i \in \mc E} \frac{c_i}{b_i}. \label{eq:lemmaineq}
\end{equation}
Then the inequality~\eqref{eq:lemmaineq} also holds for the choice $b = c$, i.e., 
\begin{equation}
\sum_{i \in \mc E} \frac{a_i}{c_i} < 1 \,.
\end{equation}
\end{proposition}

\begin{IEEEproof}
Let
\begin{equation}
i^* = \argmin_{i\in \mc E} \frac{c_i}{b_i} \,,
\end{equation}
then inequality~\eqref{eq:lemmaineq} is equivalent to 
\begin{equation}
\sum_{i \in \mc E} \frac{a_i}{b_i} \frac{b_{i^*}}{c_{i^*}} \leq 1 \,.
\end{equation}
But since $\frac{c_i^*}{b_i^*} \leq \frac{c_i}{b_i}$ for all $i \in \mc E$, it also holds that $\frac{b_i}{c_i} \leq \frac{c_{i^*}}{b_{i^*}}$ for all $i \in \mc E$ and hence
\begin{equation}
1 \geq \sum_{i \in \mc E} \frac{a_i}{b_i} \frac{b_{i^*}}{c_{i^*}} \geq \sum_{i \in \mc E} \frac{a_i}{b_i} \frac{b_{i}}{c_{i}} = \sum_{i \in \mc E} \frac{a_i}{c_i}\,.
\end{equation}
\end{IEEEproof}

\subsection{Necessary Condition for Local Networks  }
\begin{figure}
\centering
\begin{tikzpicture}[>=narrow,thick]
\node[draw,circle] (a) at (0,0) {\phantom{$v$}};
\node[draw,circle] (b) at (0,-1) {\phantom{$v$}};
\node[draw,circle] (c) at (0,1) {\phantom{$v$}};
\node[draw,circle] (d) at (2,0) {\phantom{$v$}};

\draw[->] (c) -- node[above] {$e_1$} (d);
\draw[->] (a) -- node[above] {$e_2$} (d);
\draw[->] (b) -- node[below] {$e_3$} (d);

\draw[dashed] (1, 1.5) to[bend right] (1, -1.5) node[right] {$\mc E_v$};

\end{tikzpicture}
\caption{Example of a local network, i.e., a network where all the links points towards a single node.}
\label{fig:local}

\end{figure}
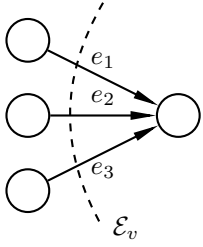

In the specific case of a local network, i.e., a network where all the links points to one node as shown in Figure~\ref{fig:local}, the condition~\eqref{eq:condcyclic} in Theorem~\ref{thm:cyclic} becomes arbitrarily close to being necessary in the case of constant inflows. For a local network, the dynamics in~\eqref{eq:generaldynamics} simplifies to 
\begin{equation}
\dot{x}_i = \lambda_i(t) - c_i f_i(x) \,,\quad \forall i \in \mc E_v\,. \label{eq:localdyn}
\end{equation}

\begin{proposition} \label{prop:localnetwork}
For a local dynamical flow network~\eqref{eq:localdyn} that satisfies Assumption~\ref{ass:flow}, if the exogenous inflows $\lambda$ are constant and 
\begin{equation}
\sum_{i \in \mc E_v} f_i(\tilde x) \leq \liminf_{\norm{x} \rightarrow +\infty} \sum_{i \in \mc E_v} f_i(x) \,, \quad \forall \tilde x \in \mc X \,, 
\end{equation}
then the condition 
\begin{equation} \label{eq:condlocal}
\sum_{i \in \mc E_v} \frac{\lambda_i}{c_i} \leq \liminf_{\norm{x}\rightarrow +\infty} \sum_{i \in \mc E_v} f_i(x)
\end{equation}
is necessary for stability of the local network~\eqref{eq:localdyn}.
\end{proposition}
\begin{IEEEproof}

Assume that 
\begin{equation}
 \sum_{i \in \mc E_v} \frac{\lambda_i}{c_i} > \liminf_{\norm{x} \rightarrow +\infty} \sum_{i \in \mc E_v} f_i(x) \,.
\end{equation}
Now observe that 
\begin{equation}
    x(t) = x(0) + \lambda t - C \int_0^t f(x(s)) \d s \,.
\end{equation}
Multiplying both sides by $\mathbf{1}^T C^{-1}$, we have
\begin{equation}
  \mathbf{1}^T C^{-1} ( x(t) - x(0)) =  \int_0^t \biggl(\sum_{i \in \mc E_v }\frac{\lambda_i}{c_i} - f_i(x(s)) \biggr) \d s \,,
\end{equation}
where the right hand side goes to infinity as $t\rightarrow +\infty$. Since $1^T C^{-1}$ will be a strictly positive vector and $x(t) > 0$ for all $t \geq 0$, it follows that $\sum_{i \in \mc E} x_i(t) \rightarrow + \infty$, which shows that condition~\eqref{eq:condlocal} is necessary. 
\end{IEEEproof}
\medskip

\begin{remark} 
In the case when the flow functions are bounded from above such that $c_i = \liminf_{x_i \rightarrow +\infty} f_i(x_i)$, the condition~\eqref{eq:condlocal} reads
\begin{equation}
\sum_{i \in \mc E_v} \frac{\lambda_i}{c_i} \leq 1 \,.
\end{equation}
This condition is similar to resource utilization conditions in processor scheduling~\cite{buttazzo2011hard} and queuing networks~\cite{bramsom}.
\end{remark}

\subsection{Alternative Bound on the Growth Rate}

While the bound~\eqref{eq:iISS} in Theorem~\ref{thm:cyclic} only provides a bound on the norm of the state, the following bound applies for each link. The bound ensures that the total amount of mass in the dynamical flow network will always be bounded by its initial state and the amount of exogenous inflow to the network and does not require Assumption~\ref{ass:flow}.

\begin{proposition}\label{prop:generalbound}
For a dynamical flow network~\eqref{eq:generaldynamics}, that satisfies Assumption~\ref{ass:routing} let $a(t) = (I -R^T)^{-1} \lambda(t)$. Then,
\begin{equation}
    x_i(t) \leq \int_0^t a_i(s) \d s + \xi_i \,, \quad \forall i \in \mc E \,, 
\end{equation}
where $\xi = (I-R^T)^{-1}x(0)$.
\end{proposition}
\begin{IEEEproof}
Let $\hat x = (I-R^T)^{-1} x$. Then
\begin{equation}
\dot{\hat x} = (I-R^T)^{-1} \lambda(t) - f(x) = a(t) - f(x)
\end{equation}
and $\hat{x}(0) = (I-R^T)^{-1}x(0)$. Since $f(x) \geq 0$, it holds that $\dot{\hat x}_i \leq a_i(t)$ for all $i \in \mc E$, and hence
\begin{equation} \label{eq:abound}
\hat x_i(t) \leq \int_0^t a_i(s) \d s + \hat{x}_i(0) \,, \quad \forall i \in \mc E\,.
\end{equation}
Observe that $\hat x (t) \geq 0$ for all $t \geq 0$ because  $(I-R^T)^{-1} = \sum_{k \geq 0} (R^T)^k$ has all elements non-negative and both $\lambda \geq 0$ and $x(0) \geq 0$. 

By transforming back to $x$, i.e., $x = (I-R^T)\hat x$, it then holds for each $i \in \mc E$ that
\begin{equation}
x_i(t) = \hat{x}_i(t) - \sum_{j \in \mc E} R_{ji} \hat{x}_j(t) \leq \hat{x}_i(t) \leq \int_0^t a_i(s) \d s + \hat{x}_i(0) \, .
\end{equation}
\end{IEEEproof}

\medskip
\begin{remark}
In~\eqref{eq:abound}, the term $\int_0^t a_i(s) \d s$  is the total mass that can possibly reach link $i \in \mc E$ from outside the network, and the term $\tilde{x}_i$ indicates how much mass can reach link $i$ from inside the network.
\end{remark}

\section{Differential Inclusion Dynamics}\label{sec:diffinc}
In the previous sections we assume that $f_i(x) = 0$ when $x_i = 0$. However, there are applications in, e.g., queuing theory and traffic signal control where this assumption does not hold. For example, in traffic signal control, it can happen that several lanes belongs to the same service phase, and then traffic present in one of the lanes will trigger the controller to serve all the lanes in the phase, including those that are empty. However, all the states will still stay non-negative.

To model this, we introduce a new state, the actual outflow $z \in \R_+^{\mc E}$. The flow dynamics now reads
\begin{equation}
\dot{x} = \lambda - (I-R^T)z \,. \label{eq:diffinc1}
\end{equation}
The outflow is non-negative and is always upper bounded by the flow functions $f(x)$, i.e.,
\begin{equation}
0 \leq z \leq f(x) \,. \retainlabel{eq:diffinc2} 
\end{equation}
Moreover, we assume that if there is mass present on the link, the amount of outflow will be equal to the one given by the flow function. This additional constraint is expressed as 
\begin{equation}
x^T(z-f(x)) = 0 \,. \label{eq:diffinc3}
\end{equation}
The dynamics~\eqref{eq:diffinc1}--\eqref{eq:diffinc3} is a differential inclusion since when $x_i = 0$ for some link $i \in \mc E$, the actual outflow $z_i$ can be anything between $0$ and $f_i(x)$.

It can be shown that the dynamics in~\eqref{eq:diffinc1}--\eqref{eq:diffinc3} is well-posted, i.e., for a given initial state $x(0)$, there exists a unique solution~\cite{como2021well-posedness}. 

In the following example, we both illustrate this differential inclusion dynamics through an example and also show why the conditions in Section~\ref{sec:stability} are not applicable in this setting.

\begin{figure}
\centering

\begin{tikzpicture}[>=narrow,thick]
\node[draw, circle] (a) at (0,-2) {\phantom{$v$}};
\node[draw, circle] (b) at (-2, 0) {\phantom{$v$}};
\node[draw, circle] (c) at (0,2) {\phantom{$v$}};
\node[draw, circle] (d) at (2, 0) {\phantom{$v$}};
\node[draw, circle] (e) at (0, 0) {\phantom{$v$}};

\draw[->] (b.-20) -- node[below] {$e_1$} (e.200);
\draw[->] (e.160) -- node[above] {$e_5$} (b.20);
\draw[->] (e.-20) -- node[below] {$e_7$} (d.200);
\draw[->] (d.160) -- node[above] {$e_3$} (e.20);

\draw[->] (c.-110) -- node[left] {$e_2$} (e.110);
\draw[->] (e.70) -- node[right] {$e_6$} (c.-70);

\draw[->] (a.70) -- node[right] {$e_4$} (e.-70);
\draw[->] (e.-110) -- node[left] {$e_8$} (a.110);
\end{tikzpicture}

\caption{The network in Example~\ref{ex:diffinc}. In this example either links ${e_1, e_3}$ or ${e_2, e_4}$ receive service simultaneously, regardless if only one of the links has mass present.}
\label{fig:onejunction}
\end{figure}
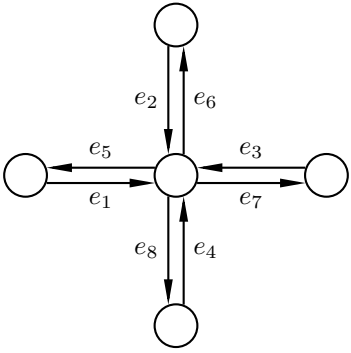

\begin{example}\label{ex:diffinc}
Consider the network in Figure~\ref{fig:onejunction} that can be seen as a model for a signalized junction in a traffic network. Suppose that either the links $e_1$ and $e_3$, or $e_2$ and $e_4$, can receive service simultaneously. In the traffic application, receiving service is equivalent to showing green light to the vehicles. Suppose now that the service is split in proportion to the total mass in each service phase, where a fraction $\kappa >0$ of the service is utilized for phase shifts. This controller is proposed and further described in~\cite{nilsson2020generalized}. Then the outflow from each link is given by
\begin{align}
f_1(x) = f_3(x) = \frac{x_1 + x_3}{x_1 + x_2 + x_3 + x_4 + \kappa} \,,\\
f_2(x) = f_4(x) = \frac{x_2 + x_4}{x_1 + x_2 + x_3 + x_4 + \kappa} \,.
\end{align}
When $x_1 = 0$ but $x_3 > 0$, $f_1(x)$ will be strictly positive although there is no mass present on the link. This justifies why the dynamical flow network dynamics has to be described through a differential inclusion as in~\eqref{eq:diffinc1}--\eqref{eq:diffinc3}.
{}
In this case,
\begin{equation}
\liminf_{\norm{x} \rightarrow +\infty} \sum_{i \in \mc E} f_i(x) = 1 + 1 = 2 \,.
\end{equation}

Hence, $\lambda_1 = 1.9, \lambda_2 = \lambda_3 = \lambda_4 = 0$ satisfies the condition in Theorem~\ref{thm:cyclic}. However, it is clear that this choice of exogenous inflows will yield an unbounded trajectory, because $f_1(x) \leq 1$ for all $x \in \mc X$.
\end{example}

\medskip

The following theorem, which generalizes the previous stated Theorem~\ref{thm:cyclic} to include non-differential inclusion dynamics, can be used to analyze dynamical flow networks with differential inclusion.

\begin{theorem}
For differential inclusion flow dynamics~\eqref{eq:diffinc1}--\eqref{eq:diffinc3} that satisfy Assumption~\ref{ass:routing} and $f(x) \geq 0$ and $f_i(x) > 0$ when $x_i > 0$, let $a(t) = (I-R^T)^{-1} \lambda$. If
\begin{equation}
\esssup_{t \geq 0} \sum_{i \in \mc E} a_i(t) < \liminf_{\norm{x} \rightarrow +\infty} \sum_{i \in \mc E} \one_{(x_i > 0)} f_i(x)
\end{equation}
then there there exists functions $\beta \in \mc K \mc L$ and $\mu \in \mc K_\infty$ such that every continuous solution to the dynamics~\eqref{eq:diffinc1}--\eqref{eq:diffinc3} satisfies
\begin{multline}
    \norm{(I-R^T)^{-1}x(t)} \leq \beta(\norm{(I-R^T)^{-1} x(0)}, t)  + \\ \mu\left(\esssup_{t \geq 0} \norm{(I-R^T)^{-1}\lambda(t)}\right) \,.
\end{multline}
\end{theorem}

\begin{IEEEproof}
Introduce the Lyapunov candidate
\begin{equation}
V(x) = \1^T(I-R^T)^{-1}x \,.
\end{equation}
Since $(I-R^T)^{-1} = I + R^T + (R^T)^2 + \ldots$ and $x\geq 0$, it holds that $V(x) \geq 0$ and $V(x) = 0$ if and only if $x=0$. Moreover,
\begin{align}
\frac{\d V}{\d t} &= \1^T(I-R^T)^{-1}\dot{x} \\
&= \sum_{i \in \mc E} a_i(t) - \sum_{i \in \mc E : x_i > 0} f_i(x) - \sum_{i \in \mc E : x_i = 0} z_i \\
&\leq  \sum_{i \in \mc E} a_i(t) + \sum_{i \in \mc E } \one_{(x_i> 0)} f_i(x) \,. 
\end{align}
Now, following the same methodology as in the proof of Theorem~\ref{thm:cyclic}, but with  $W(x) = \sum_{i \in \mc E } \one_{(x_i> 0)} f_i(x)$ instead (which is positive definite, since $f_i(x)$ is assumed to be strictly positive when $x_i > 0$), \cite[Theorem 1]{strongiISS} can again be applied to obtain the result.
\end{IEEEproof}

\section{Numerical Examples}\label{sec:examples}
In this section we present two examples, both illustrative how the theory in this paper allows to analyze the stability of dynamical flow networks for broader classes of dynamics than was previously possible.

\subsection{Time-Varying Inflows} \label{sec:timevarying}

\begin{figure}
\centering
\begin{tikzpicture}[>=narrow,thick]
\node[draw, circle] (a) at (0,0) {$v_1$};
\node[draw, circle] (b) at (2,0) {$v_2$};
\node[draw, circle] (c) at (4,0) {$v_3$};
\draw[->] (a) to[bend left] node[above] {$e_1$} (b);
\draw[->] (a) to[bend right] node[below] {$e_2$} (b);
\draw[->] (b) to[bend left] node[above] {$e_3$} (c);
\draw[->] (b) to[bend right] node[below] {$e_4$} (c);
\end{tikzpicture}
\caption{The network for the example in Section~\ref{sec:timevarying}.}
\label{fig:networktwostage}
\end{figure}
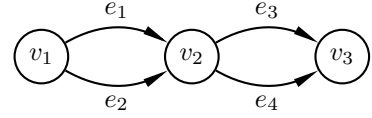

Consider the network in  Figure~\ref{fig:networktwostage}. Suppose that there are only exogenous inflows to the first two nodes, i.e., $\lambda_3 = \lambda_4 = 0$ and let $R_{1,4} = R_{1,3} = 0.5$ and $R_{2,4} = 1$. Suppose that each node splits the outflow from the incoming links according to
\begin{equation}
    f_i (x) = \frac{x_i}{\sum_{j \in \mc E_v} x_j + 1} \,, \quad \forall i \in \mc E_v\,, \forall v \in \mc V =\{v_2, v_3\} \,.
\end{equation}
For static inflows, the sufficient and necessary condition for bounded state presented in~\cite{nilsson2020generalized} is $\lambda_1 + \lambda_2 < 1$.

\begin{figure}
    \centering
    \input{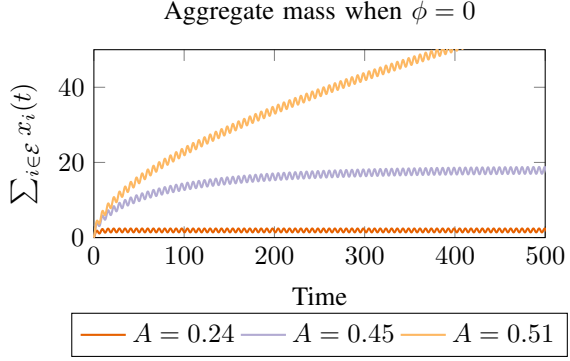}
    \caption{The aggregate mass on all the links for the example in Section~~\ref{sec:timevarying}. Although the sufficient condition only ensures stability for $A = 0.24$, the system is stable for $A = 0.45$ too. For $A = 0.51$, the trajectory diverges.}
    \label{fig:twostageplots}
\end{figure}

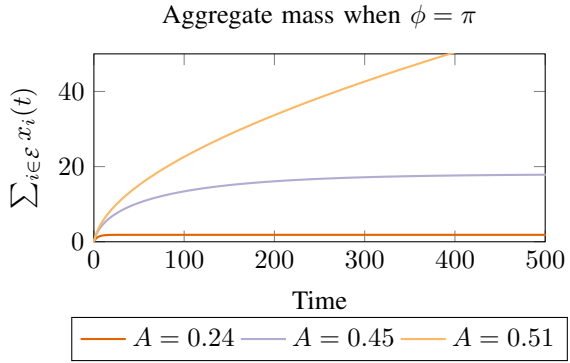
\begin{figure}
    \centering
    \begin{tikzpicture}

\begin{axis}[%
width=6cm,
height=2.5cm,
scale only axis,
xmin=0,
xmax=500,
ymin=0,
ymax=50,
xlabel={Time},
ylabel={$\sum_{i \in \mc E} x_i(t)$},
ylabel near ticks,
legend style={at={(0.5,-0.4)},anchor=north},
legend columns=3,
title={Aggregate mass when $\phi = \pi$}
]
\addplot [color=mycolor1, thick]
  table[row sep=crcr]{%
0	0\\
0.210000000000036	0.100243260878869\\
0.330000000000041	0.156303648040307\\
0.440000000000055	0.206486585524203\\
0.539999999999964	0.250952347023599\\
0.639999999999986	0.29423450369768\\
0.740000000000009	0.336288846948264\\
0.840000000000032	0.377097170263369\\
0.940000000000055	0.416660077155939\\
1.03999999999996	0.454991573754228\\
1.13999999999999	0.492115030016294\\
1.24000000000001	0.528060179592217\\
1.34000000000003	0.562860902434295\\
1.45000000000005	0.599863367993862\\
1.55999999999995	0.635575376035604\\
1.66999999999996	0.670047741708458\\
1.77999999999997	0.703330911368994\\
1.88999999999999	0.735474315556985\\
2.00999999999999	0.769296292509125\\
2.13	0.801878895938103\\
2.25	0.833279432423637\\
2.37	0.863552443907906\\
2.5	0.89513579257607\\
2.63	0.925518754660743\\
2.75999999999999	0.954759874427054\\
2.89999999999998	0.985036423553765\\
3.03999999999996	1.01411701673408\\
3.17999999999995	1.04206199752196\\
3.33000000000004	1.07080694127501\\
3.48000000000002	1.09837787913671\\
3.63	1.1248352868646\\
3.78999999999996	1.15189267618473\\
3.95000000000005	1.17781110551164\\
4.12	1.2041678640212\\
4.28999999999996	1.22937215764341\\
4.47000000000003	1.25487128960629\\
4.64999999999998	1.27921528844763\\
4.84000000000003	1.30372811500297\\
5.02999999999997	1.32709281580355\\
5.23000000000002	1.35051690374564\\
5.42999999999995	1.37280838351626\\
5.63999999999999	1.39506533484803\\
5.86000000000001	1.41719301364935\\
6.08000000000004	1.43817568575548\\
6.30999999999995	1.45896104587632\\
6.54999999999995	1.47947156437806\\
6.78999999999996	1.49885225413936\\
7.03999999999996	1.51791361193727\\
7.29999999999995	1.53659424136322\\
7.57000000000005	1.55483900245781\\
7.85000000000002	1.57259898125835\\
8.13999999999999	1.58983140492114\\
8.44000000000005	1.60649950964068\\
8.75	1.62257236805897\\
9.07000000000005	1.63802468228278\\
9.40999999999997	1.65326849065821\\
9.75999999999999	1.66779287535496\\
10.12	1.68159283972261\\
10.5	1.69500653830846\\
10.9	1.70795328534712\\
11.3099999999999	1.72008249867406\\
11.74	1.73167239139048\\
12.2	1.74290728986716\\
12.6799999999999	1.75347204572563\\
13.1900000000001	1.76353206838701\\
13.73	1.77301459289981\\
14.3	1.78186511881404\\
14.91	1.79017406526702\\
15.5599999999999	1.79787026979841\\
16.26	1.80500065539229\\
17.02	1.81157657951599\\
17.85	1.81758525145381\\
18.76	1.82300098413657\\
19.77	1.82783867928345\\
20.9	1.83208255009527\\
22.1900000000001	1.83575688124449\\
23.6900000000001	1.83885715665394\\
25.49	1.84139722084899\\
27.74	1.84337940872911\\
30.74	1.84481066632304\\
35.24	1.84570690516057\\
44.05	1.84610371288773\\
93.62	1.84615384600443\\
999.99	1.84615384615381\\
};
\addlegendentry{$A = 0.24$}

\addplot [color=mycolor2, thick]
  table[row sep=crcr]{%
0	0\\
0.409999999999968	0.362247771747889\\
0.67999999999995	0.58647903146209\\
0.950000000000045	0.797200276285366\\
1.23000000000002	1.00220235945199\\
1.51999999999998	1.20145777616437\\
1.82000000000005	1.39524465011868\\
2.13999999999999	1.58986002902213\\
2.48000000000002	1.78472783898587\\
2.85000000000002	1.98467683478611\\
3.24000000000001	2.18356371038851\\
3.65999999999997	2.38591288370685\\
4.10000000000002	2.58642883738582\\
4.57000000000005	2.78929802129142\\
5.07000000000005	2.99387745918762\\
5.60000000000002	3.19962820513535\\
6.16999999999996	3.4096909031623\\
6.76999999999998	3.6197381845252\\
7.40999999999997	3.83273225275229\\
8.09000000000003	4.04796548380898\\
8.80999999999995	4.26482715984798\\
9.57000000000005	4.48279021898429\\
10.37	4.70139950282339\\
11.21	4.92026150207062\\
12.1	5.14146408583258\\
13.04	5.36438195321148\\
14.03	5.58846548853489\\
15.0700000000001	5.81323178881826\\
16.16	6.03825654236221\\
17.3099999999999	6.26509550743231\\
18.52	6.49316592496882\\
19.79	6.72195035068512\\
21.12	6.95098965981049\\
22.51	7.17987662811606\\
23.97	7.40979047366034\\
25.5	7.64021053793851\\
27.1	7.87067150049961\\
28.77	8.10075786635662\\
30.52	8.3313880477117\\
32.35	8.56206382374523\\
34.26	8.79233802861734\\
36.25	9.02180980857213\\
38.33	9.25119886681273\\
40.49	9.47901154748081\\
42.75	9.7069400597278\\
45.1	9.93353228855017\\
47.55	10.1593545600766\\
50.1	10.3839874258381\\
52.75	10.6070533222052\\
55.51	10.8289990701911\\
58.38	11.049414575619\\
61.37	11.2686469399421\\
64.48	11.4862661384429\\
67.71	11.7018844979883\\
71.0600000000001	11.9151535318628\\
74.54	12.1263534736878\\
78.16	12.3356859041776\\
81.92	12.5427444943491\\
85.8200000000001	12.7471640181305\\
89.87	12.9491038014239\\
94.08	13.1486624195965\\
98.45	13.3454454659259\\
102.99	13.5395165546922\\
107.7	13.7304972195659\\
112.59	13.9184245494671\\
117.67	14.1032946711372\\
122.94	14.2847331763049\\
128.42	14.4630414164679\\
134.11	14.6378221257776\\
140.02	14.8090060880536\\
146.16	14.9765023999983\\
152.54	15.1402025556549\\
159.18	15.3002177705471\\
166.09	15.4563716920242\\
173.28	15.6084849521334\\
180.76	15.7563784104357\\
188.56	15.900232279041\\
196.69	16.0398021852975\\
205.17	16.1750128210962\\
214.02	16.3057611339633\\
223.27	16.4320549467897\\
232.95	16.5538471472369\\
243.08	16.6709367506018\\
253.71	16.7834324128877\\
264.87	16.89116239192\\
276.61	16.9941142344767\\
288.98	17.0922126691104\\
302.04	17.1854057870613\\
315.87	17.2737068844968\\
330.55	17.3570407576101\\
346.18	17.4353672056817\\
362.88	17.5086495246805\\
380.8	17.5768699295363\\
400.12	17.6399949646561\\
421.07	17.6980087557821\\
443.94	17.7508894612527\\
469.11	17.7986243140042\\
497.1	17.8412211972591\\
528.61	17.878665919828\\
564.68	17.9109841251458\\
606.86	17.9381843878926\\
657.68	17.9602978287398\\
721.67	17.9773773170333\\
808.1	17.989501020269\\
941.08	17.9968268030017\\
999.99	17.998141873053\\
};

\addlegendentry{$A = 0.45$}
\addplot [color=mycolor3, thick]
  table[row sep=crcr]{%
0	0\\
0.75	0.729187261791139\\
1.37	1.25325388216925\\
2.07000000000005	1.7735996491474\\
2.87	2.3007665162346\\
3.78999999999996	2.84277561796864\\
4.85000000000002	3.40522245005263\\
6.07000000000005	3.99189823371762\\
7.46000000000004	4.60112714024478\\
9.03999999999996	5.23557330268079\\
10.83	5.89703038367395\\
12.84	6.58339551563915\\
15.1	7.2991830777587\\
17.62	8.04193048332229\\
20.4299999999999	8.81504956273034\\
23.54	9.61605144246437\\
26.98	10.4476622042737\\
30.77	11.309723034048\\
34.9299999999999	12.2020570330009\\
39.48	13.1244828317404\\
44.46	14.0805640598081\\
49.88	15.0677952579383\\
55.77	16.0875036722557\\
62.16	17.1407513908339\\
69.08	18.2283924300299\\
76.55	19.3496499004972\\
84.61	20.5066814200917\\
93.28	21.6985919000052\\
102.6	22.927212880971\\
112.6	24.1928526498148\\
123.31	25.495778132238\\
134.76	26.8362261183064\\
146.99	28.2155196953947\\
160.03	29.6337436237907\\
173.91	31.0909886126508\\
188.68	32.5893492019399\\
204.37	34.1287411586399\\
221.02	35.7100494221359\\
238.66	37.3331704681949\\
257.34	38.9997989836121\\
277.1	40.7106086818359\\
297.98	42.4662276483183\\
320.02	44.267246849454\\
343.26	46.1142273185072\\
367.75	48.0084695198279\\
393.53	49.9504328422751\\
420.65	51.9412930040698\\
449.16	53.9821489765735\\
479.1	56.0733439193129\\
510.52	58.2159034894951\\
543.47	60.4108010522249\\
578	62.6589653460223\\
614.16	64.9612871510155\\
652	67.3186250937747\\
691.58	69.7324142314754\\
732.95	72.2034265555186\\
776.16	74.732418347207\\
821.27	77.3207014564887\\
868.33	79.9689771300077\\
917.41	82.679035720209\\
968.56	85.4515105741914\\
999.99	87.1306387596561\\
};
\addlegendentry{$A = 0.51$}

\end{axis}
\end{tikzpicture}%
    \caption{The aggregate mass on all the links for the example in Section~~\ref{sec:timevarying}. In this case, with $\phi = \pi$, the sufficient condition ensures stability for both $A = 0.24$ and $A = 0.45$. For $A = 0.51$ the condition is not satisfied and the trajectory diverges.}
    \label{fig:twostageplots2}
\end{figure}

Theorem~\ref{thm:cyclic} makes it possible to ensure stability for time-varying inflows such that the sufficient condition to ensure bounded state is now 
\begin{equation}
    \esssup_{t \geq 0} \lambda_1(t) + \lambda_2(t) < 1 \,. \label{eq:condtwonodes}
\end{equation}
However, this condition generally only sufficient. To illustrate this, we let 
\begin{equation}
\lambda_1(t) = A(\sin(t) +1) \,, \quad
\lambda_2(t) = A(\sin(t + \phi) + 1) \,,
\end{equation}
and we consider 
several choices of $A$ and $\phi$. When $\phi = 0$, the sufficient condition in~\eqref{eq:condtwonodes} is equivalent to $A < 0.25$. However, as can it can be seen in Figure~\ref{fig:twostageplots}, the state remains bounded for $A = 0.45$ but becomes unbounded for $A=0.51$, which illustrates the sufficiency of the condition. By instead letting $\phi = \pi$, the sufficient condition in~\eqref{eq:condtwonodes} is now equivalent to $A < 0.5$.  The aggregate mass trajectories are shown in Figure~\ref{fig:twostageplots2} for the different choices of $A$ when $\phi = \pi$. 

\subsection{Multi-Commodity Flows}\label{sec:multicommodity}
Although the analysis in this note is done for single commodity flows, it can be extended to multi-commodity flows, i.e., dynamical flow networks where different commodities share the same network, but differ in routing. In particular, the technique can be used to accommodate any finite number of commodities, but for the purpose of this example, assume that we have two commodities, which we will denote $A$ and $B$. Let $\lambda^A, \lambda^B \in \R_{+}^{\mc E}$ denote the exogenous inflow of the respective commodity. Each commodity is assumed to have its own routing matrix, which we denote $R^A$ and $R^B$, where both routing matrices are assumed to satisfy Assumption~\ref{ass:routing} individually. As state space, we now need to keep track of the mass of each commodity on every link in the network, i.e., the state is $(x^A, x^B)$ with $x^A, x^B \in \R_+^{\mc E}$. We let the vector $x \in \R_+^{\mc E}$ be the aggregate mass on each link, i.e., $x_i = x_i^A + x_i^B$ for every link $i \in \mc E$. Under the assumption that the commodities are perfectly mixed and travel with the same aggregate flow dynamics, the dynamics for the flow network with bounded outflow functions becomes
\begin{align}
 \dot{x}^A &= \lambda^A - (I-(R^A)^T)C \text{diag}\left(\left(\frac{x_i^A}{x_i}\right)_{i \in \mc E}\right) f(x) \,, \\
\dot{x}^B &= \lambda^B - (I-(R^B)^T)C \text{diag}\left(\left(\frac{x_i^B}{x_i}\right)_{i \in \mc E}\right) f(x) \,.
\end{align}
This model has previously been used to study road traffic flows where different commodities have different routing objectives in~\cite{nilsson2018twotier}. Different from the stability results presented in that paper, here, we allow the network to contain cycles.

\begin{figure}
\centering 
\begin{tikzpicture}[>=narrow, thick]
    
\node[draw, circle] (0) at (0,0) {\phantom{$v$}};
\node[draw, circle] (1) at (2,1) {\phantom{$v$}};
\node[draw, circle] (2) at (2,-1) {\phantom{$v$}};
\node[draw, circle] (3) at (4,0) {\phantom{$v$}};
    
\draw[->] (-2, 0)  -- node[below] {$e_1$} (0);
\draw[->] (0) --  node[below] {$e_2$} (1);
\draw[->] (1) --  node[left] {$e_3$} (2);
\draw[->] (1) --  node[below] {$e_4$} (3);
\draw[->] (0) --  node[below] {$e_5$} (2);
\draw[->] (2) --  node[below] {$e_6$} (3);
\draw[->] (1) to[bend right]  node[above] {$e_7$} (0);    
\end{tikzpicture}

\caption{The multi-commodity flow network used in the example in Section~\ref{sec:multicommodity}.}
\label{fig:multicommnetwork}
\end{figure}
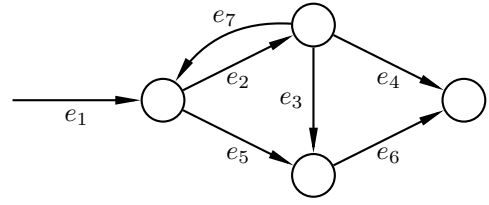

Consider the network in Figure~\ref{fig:multicommnetwork}. Let the outflow functions for each link be $f_i(x_i) = 6 (1-e^{- x_i})$. The non-zero elements in the routing matrix for each commodity are specified in Table~\ref{tab:multicommrouting}. 

\begin{table} 
    \centering
        \caption{The non-zero elements in the routing matrices} \label{tab:routing}
    \begin{tabular}{c|cc}
         & Commodity A & Commodity B \\ \hline
    $R_{1,2}$ & $0.6$ & $0.7$ \\
    $R_{1,5}$ & $0.4$ & $0.3$\\
    $R_{2,7}$ & $0.1$ & $0.3$\\
    $R_{2,3}$ & $0.3$ & $0.4$\\
    $R_{2,4}$ & $0.6$ & $0.3$\\
    $R_{3,6}$ & $1$ &  $1$\\
    $R_{5,6}$ & $1$ & $1$ \\ 
    $R_{7,2}$ & $0.5$ & $0.3$\\ 
    $R_{7,5}$ & $0.5$ & $0.7$\\\hline
    \end{tabular}
    \label{tab:multicommrouting}
\end{table}

Define
\begin{align}
    a^A(t) = (I-(R^A)^T)^{-1} \lambda^A(t) \,, \\
    a^B(t) = (I-(R^B)^T)^{-1} \lambda^B(t) \,.
\end{align}
By using the Lyapunov function
\begin{equation}
    V(x) = \1^T C^{-1}(I-(R^A)^T)^{-1} x^A + \1^T C^{-1}(I-(R^B)^T)^{-1}x^B \,,
\end{equation}
and the same theory as in the proof of Theorem~\ref{thm:cyclic} and~\ref{thm:cyclicbounded}, we obtain the following sufficient condition for stability of the multi-commodity dynamics:
\begin{equation} \label{eq:condmulticommex}
\esssup_{t \geq 0} \sum_{i \in \mc E} \frac{a^A_i(t) + a^B_i(t)}{c_i} < \liminf_{\norm{x} \rightarrow +\infty} \sum_{i \in \mc E} f_i(x) \,.
\end{equation}

\begin{figure}
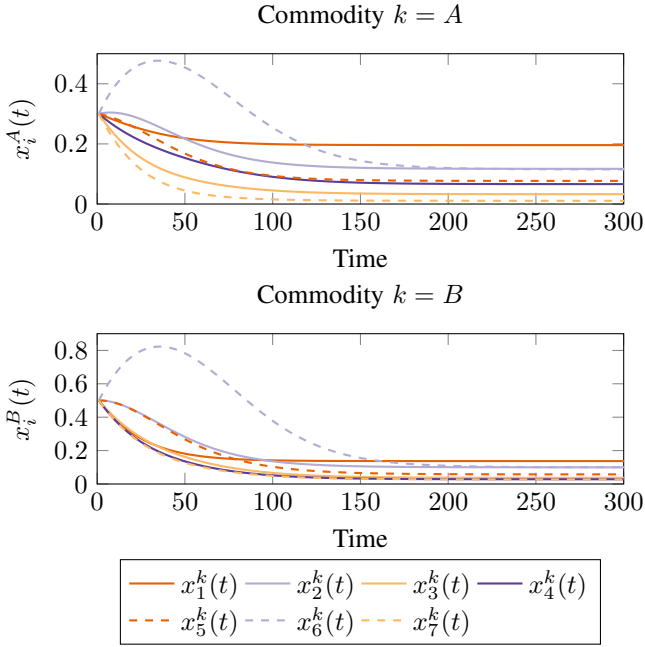

    \centering
    \input{figures/multicommA.tikz} 
    \input{figures/multicommB.tikz}
    \caption{The trajectories for commodity $A$ and commodity $B$ respectively in the example in Section~\ref{sec:multicommodity}.}
    \label{fig:multicommA}
\end{figure}

In this example, note that $\liminf_{\norm{x} \rightarrow +\infty} \sum_{i \in \mc E} f_i(x_i) = 1$. If we let $\lambda^A_1 = 1$, $\lambda^B_1 = 0.7$ and all other elements of $\lambda^A, \lambda^B$ be zero, the sufficient condition in~\eqref{eq:condmulticommex} is satisfied. The trajectory for each commodity is shown in Figure~\ref{fig:multicommA}, with the initial state $x^A_i(0) = 0.3$ and $x^B_i(0) = 0.5$ for all $i \in \mc E$.

\section{Conclusions}

In this note, we have shown how the Strong integral Input-to-State Stability (Strong iISS) framework is particularly suitable for studying  the stability of dynamical flow networks. We established sufficient conditions on the exogenous inflows for dynamical flow networks to be stable, i.e., ensure that the state remains bounded, and showed that the condition is also necessary for certain types of networks. We also showed how the conditions can be applied to existing dynamical flow network models and provided stability assurance in settings not covered in prior literature.

A future research direction is to explore if the theory of Strong iISS leads to alternative tighter bounds in certain settings, e.g., by considering the time-averaged exogenous inflow or dividing the network into several sub-networks. 

\bibliographystyle{IEEEtran}
\bibliography{references.bib} 
\end{document}